\documentclass[12pt]{article}
\usepackage[latin1]{inputenc}
\usepackage{latexsym}
\usepackage{amstext}
\usepackage{amsxtra}
\usepackage{epsf}

\title{A Monte Carlo Formulation of the Bogolubov Theory}
\author{ Alice Sinatra and Yvan Castin \\  
{\small Laboratoire Kastler Brossel, 24 Rue Lhomond, 75231 Paris Cedex 05, France}\\
Carlos Lobo \\
{\small University of Illinois at Urbana-Champaign, 1110 West Green Street, Urbana, 
Illinois 61801-3080} }

\begin{document}
\maketitle
\begin{abstract}
We propose an efficient stochastic method to implement numerically the Bogolubov
approach to study finite-temperature Bose-Einstein condensates. 
Our method is based on the Wigner representation of the density matrix describing the
non condensed modes and a Brownian motion simulation to sample the Wigner
distribution at thermal equilibrium. Allowing to sample any density operator Gaussian
in the field variables, our method is very general and it applies both to the Bogolubov 
and to the Hartree-Fock Bogolubov approach, in the equilibrium case as well as in the 
time-dependent case. 
We think that our method can be useful to study trapped Bose-Einstein condensates 
in two or three spatial dimensions without rotational symmetry properties, 
as in the case of condensates with vortices, where the traditional Bogolubov approach 
is difficult to implement numerically due to the need to diagonalize very big matrices.
\end{abstract}

\section{Introduction}
Since the first experimental demonstrations of Bose-Einstein condensation in atomic 
gases \cite{BEC1,BEC2,BEC3,BEC4}, we have assisted to a fantastic development 
of the field both on experimental and theoretical ground. From the theoretical point of view
the Gross-Pitaevskii equation (GPE) or non-linear Schr\"odinger equation for the condensate wave 
function, in its steady-state and time dependent versions, 
revealed itself to be an extremely powerful tool to explain most of the experimental 
results \cite{rev1}. 

Generally speaking the Gross-Pitaevskii equation is obtained by neglecting 
the interaction between the condensate and non condensed particles \cite{Pines}.
This approximation is justified at temperatures much below the transition temperature 
where the non condensed cloud is very small and much less dense that the condensate. This 
is often the case in experimental conditions where almost pure condensates can be obtained. 

There are nevertheless situations in which the rich physics coming from the non condensed 
fraction of the cloud and its coupling to the condensate cannot be neglected.
A first example concerns the collective modes of the gas induced by a modulation of the 
trapping potential \cite{JILAcoll,MITcoll}.
Different theories going beyond the Gross-Pitaevskii equation were put forward
in order to explain the temperature dependence of the frequencies and the damping rates 
measured in experiments \cite{Griffin,Shlyap,Stoof} (for more references 
see \cite{rev1}, pages 489-500).

A second example where the non condensed fraction may play a significant role
concerns condensates in thermodynamically unstable states, like vortices
and dark solitons recently created experimentally \cite{JILAvortex,ENSvortex,Hannover,NIST}.
In \cite{ENSvortex} for example one or more vortices are created by stirring the condensate with a slightly
anisotropic rotating potential: vortices appear provided that the rotation frequency exceeds a critical 
value. If the rotation is then stopped the vortices are not any more a local minimum of energy
and are expected to spiral out of the condensate \cite{Rockstar,Yvanvortex}. It is indeed found 
experimentally that the vortices leave the condensate \cite{ENSvortex}.
It is predicted theoretically that the dynamics and lifetime of the thermodynamically 
unstable Bose-Einstein condensate depend crucially on the interaction between the condensate 
and the non condensed cloud \cite{Giora}. 

The examples we have given above concern time dependent properties of the condensate. 
Another class of interesting finite temperature issues is given by the thermal 
equilibrium properties of the gas.  
To cite some example, one might be interested in the non condensed spatial density 
of the atoms, or in the thermal fluctuations of the number of particles in the condensate
\cite{Stringarifluc,Wilkens,Scully,Houches}. 
Condensates with vortices give rise again to interesting questions
in this respect such as the thermal fluctuations in the position of the vortex core and 
the temperature dependence of the critical rotation velocity \cite{Stringari}. 

A well established technique to study Bose-Einstein condensates at thermal equilibrium 
is the Bogolubov method, which is a perturbative method valid for low temperatures when the density 
of condensate particles is much larger than the density of non condensed particles. This method   
relies on the quadratization of the Hamiltonian in the field operator representing the
non condensed fraction, and on the Bogolubov transform \cite{Bogol}. 
Another method which could be valid for a larger range of temperatures is the 
Hartree-Fock-Bogolubov (HFB) method \cite{HFB,GriffinHFB}, 
which is instead a variational method, based on a Gaussian
Ansatz for the system density operator.
For time dependent problems, one disposes of the former two methods, plus more 
refined techniques including 
kinetic equations for the non condensed particles \cite{Gardiner,Kagan,GriffinCin,StoofCin}.

A common point to all these methods is the necessity to calculate mean values using a density
operator which is an exponential of a quadratic form in the field operator:
the exponential of the quadratized Hamiltonian in the case of the Bogolubov approach or the 
Gaussian Ansatz for the HFB approach.  
The procedure that is usually followed to perform those calculations is to diagonalize the 
quadratic form by the Bogolubov transform, turning it into a sum of decoupled harmonic 
oscillator energy terms.

Unfortunately the corresponding calculations
are quite heavy for 3D non-homogeneous systems such as trapped gases,
and this constitutes a serious limitation to the use of these methods.  
E.g.\ in practice, by discretizing the real space on a grid with a modest number of
$64$ points per spatial dimension, and in the absence of rotational symmetry (as in a
multi-vortex configuration)
one has to diagonalize a matrix which is $[2 (64)^n \times 2 (64)^n]$ 
large, where $n=1,2,3$ is the dimension of the space. For $n=3$ the matrix to diagonalize
is $524\, 288\times 524\, 288$.

In this paper we propose an alternative formulation of the Bogolubov theory which can be
implemented numerically more efficiently in two or three spatial dimensions since it avoids
the diagonalization of {\sl big} matrices. Our method relies on the use of the Wigner 
representation of the density  operator
and on a Monte Carlo simulation to sample the steady-state Wigner distribution
of the system at finite temperature.
Since our method allows to sample a ``Gaussian" density operator,
it applies both to the Bogolubov and to the HFB theory,
in the thermal equilibrium as well as in the time dependent case. 
We give in this paper a detailed derivation of the method in the frame of a particle number
conserving Bogolubov theory \cite{Girardeau,GardinerBogol,CastinDum}. 
We consider first the thermal equilibrium case and then we show how to extend the treatment
to the time dependent case. 

The paper is organized as follows: in section \ref{sec:bogol} we recall the number-conserving version
of the Bogolubov theory put forward in \cite{CastinDum}, which we adopt in the following.
In section \ref{sec:wigner} we reformulate the theory by expressing the Bogolubov density operator 
in terms of the Wigner quasi-distribution function. In section \ref{sec:simul} we describe the
stochastic simulation that we use to sample the equilibrium Wigner function.
In section \ref{sec:resul} 
we compare numerically our stochastic method to the direct diagonalization used in the traditional
Bogolubov approach in the case of a 1D trapped Bose gas, where the explicit diagonalization is
feasible, and we calculate the spatial distribution of the non condensed fraction.  
In section \ref{sec:timedep} we show how a time dependent problem would be treated with 
our method. Conclusions and perspectives are presented in section \ref{sec:concl}.

\section{Bogolubov theory conserving the total number of particles}
\label{sec:bogol}

In this section we recall briefly the main ideas and results of \cite{CastinDum} 
where a number-conserving version of the Bogolubov theory is established.
The reader already familiar with this theory can jump directly to section \ref{sec:wigner}.

We consider a cloud of atoms trapped in the potential $U(\vec{r}\,)$
interacting through an effective low energy contact potential 
$V= g \delta(\vec{r_1}-\vec{r_2}\, )$:
\begin{equation}
H =\int d^3{\vec{r}} \; \hat{\psi}^\dagger(\vec{r}\,) h_1 \hat{\psi}(\vec{r}\,)
+ \frac{g}{2} \hat{\psi}^\dagger(\vec{r}\,)\hat{\psi}^\dagger(\vec{r}\,)
  \hat{\psi}(\vec{r}\,)\hat{\psi}(\vec{r}\,)
\end{equation}
where $\hat{\psi}$ is the atomic field operator, $h_1$ is the one-body Hamiltonian:
\begin{equation}
h_1 = -\frac{\hbar^2}{2m}\Delta + U(\vec{r}\,) \,.
\end{equation}
The coupling constant $g$ is expressed in terms of the $s$-wave scattering
length $a$ of the true interaction potential:
\begin{equation}
{ g} = \frac{4\pi\hbar^2{ a}}{m} \;.
\end{equation}

Let $\rho_1$ be the one-body density operator of our system
and let $\phi_{\rm ex}$ be the eigenvector of $\rho_1$ with the largest eigenvalue $N_0$:  
\begin{equation}
\rho_1|\phi_{\rm ex}\rangle = N_0 |\phi_{\rm ex}\rangle \,.
\label{eq:auto}
\end{equation}
We normalize $\phi_{\rm ex}$ to unity.
Physically $\phi_{\rm ex}$ is the most populated single particle state. We assume here the 
presence of a Bose-Einstein condensate so that $\phi_{\rm ex}$ is the condensate wave
function and $N_0 \simeq N$.
This motivates the splitting of the atomic field operator $\hat{\psi}$ 
into condensate and non condensed modes:
\begin{equation}
\hat{\psi}(\vec{r}\,)=\phi_{\rm ex}(\vec{r}\,)\hat{a}_{\phi_{\rm ex}} + \delta\hat{\psi}(\vec{r}\,)
\label{eq:psi}
\end{equation}
where $\hat{a}_{\phi_{\rm ex}}$ annihilates a particle in the condensate wavefunction $\phi_{\rm ex}$, and
$ \delta\hat{\psi}(\vec{r}\,)$ is orthogonal to $\phi_{\rm ex}$:
\begin{equation}
\int \phi_{\rm ex}^* \delta\hat{\psi} = 0 \,.
\label{eq:ortho}
\end{equation}
An important property that comes from (\ref{eq:auto}) and (\ref{eq:ortho}) it that there 
is no single particle coherence between the condensate and the non condensed modes:
\begin{equation}
\langle \hat{a}^\dagger_{\phi_{\rm ex}} \delta\hat{\psi} \rangle = 0 \,.
\label{eq:nocoh}
\end{equation}
The interest of the splitting (\ref{eq:psi}) is that
the typical matrix element of $\hat{a}_{\phi_{\rm ex}}$ is of order $N_0^{1/2}\simeq N^{1/2}$ while 
the matrix elements of $\delta \hat{\psi}$ scale as $(\delta N)^{1/2} \ll N^{1/2}$ where $\delta N$ is
the number of non condensed particles. 

An essential ingredient of the number conserving Bogolubov approach is 
the operator $\hat{\Lambda}_{\rm ex}$ transferring one non condensed particle into the condensate:
\begin{equation}
\hat{\Lambda}_{\rm ex}= \hat{N}^{-1/2} \,
 \hat{a}_{\phi_{\rm ex}}^\dagger \, \delta\hat{\psi} \;.
\end{equation}
This operator plays here the role played by 
$\delta \hat{\psi}$ in the symmetry breaking approach commonly adopted.
In particular due to (\ref{eq:nocoh}), the expectation value of 
$\hat{\Lambda}_{\rm ex}$ vanishes.

In the weakly interacting Bose gas limit 
\begin{equation}
N \rightarrow \infty \hspace{2cm} \mbox{and} \hspace{2cm} g \;N = {\rm const}\,, 
\label{eq:tdlimit}
\end{equation}
and for a constant trapping potential,
the mean interaction energy per particle $n\,g$ (where $n$ is the density) tends to a 
finite value whereas the small gaseous parameter $(n a^3)^{1/2}$ tends to zero.
If the temperature, the trapping potential and the mean interaction energy are fixed,
the number of non condensed particles $\delta N$ is bounded from above so that $\delta N \ll N$
in the large $N$ limit \cite{rev1}.
One can then make a systematic expansion of the exact condensate wavefunction $\phi_{\rm ex}$
and of the fields $\hat{\Lambda}_{\rm ex}$ and $\hat{\Lambda}_{\rm ex}^\dagger$ in powers of the 
small parameter:
\begin{equation}
\epsilon=\sqrt{\delta N/N} \sim 1/\sqrt{N} \,. 
\end{equation}
Formally:
\begin{eqnarray}
\hat{\Lambda}_{\rm ex} &=& \hat{\Lambda}_{(0)} + \frac{1}{\sqrt{N}}  
	\hat{\Lambda}_{(1)} + \frac{1}{N} \hat{\Lambda}_{(2)} + \ldots 
	 \\
\phi_{\rm ex} &=& \phi_{(0)} + \frac{1}{\sqrt{N}}  
	\phi_{(1)} + \frac{1}{N} \phi_{(2)} + \ldots 
\end{eqnarray}
One then calculates  $ {d} \langle \hat{\Lambda}_{\rm ex} \rangle / dt = 0 $
order by order in ${ \epsilon}$. At the order $-1$ in $\epsilon$ (the lowest approximation
order), one finds $\hat{\Lambda}_{(0)}=0$ and $\phi_{(0)}=\phi$ where $\phi$
is solution of the time-dependent Gross-Pitaevskii equation:  
\begin{equation}
i \hbar \partial_t \phi = [ h_1 + Ng |\phi|^2 - \mu] \, \,\phi \;.
\label{eq:GPE}
\end{equation}
The quantity $\mu$ in (\ref{eq:GPE}) is such that the wavefunction
of the condensate at equilibrium is a time independent solution of the above equation. 
Physically $\mu$ is the lowest order approximation to the chemical potential of the gas.

The order $0$ in $\epsilon$ provides the equations of motion for  
$\hat{\Lambda}\equiv \frac{1}{\sqrt{N}} \hat{\Lambda}_{(1)}$ and $\hat{\Lambda}^\dagger$: 
\begin{equation}
i \hbar \partial_t   \left(
\begin{tabular}{c} ${\hat{\Lambda}}$ \\ $\hat{\Lambda}^\dagger$
\end{tabular}\right) = {\cal L}
\left(
\begin{tabular}{c} ${\hat{\Lambda}}$ \\ ${\hat{\Lambda}}^\dagger$
\end{tabular}\right) \;.
\label{eq:evollambda}
\end{equation}
The first non zero correction to $\phi_{(0)}$ is $\frac{1}{N} \phi_{(2)}$, whose equation
of motion is obtained in the next order of the expansion. 
The operator ${\cal L}$ in (\ref{eq:evollambda}) is given by:
\begin{center}
\begin{equation}
{ {\cal L}}=\left( \begin{tabular}{cc}
$H_{\rm gp}-\mu + { Q} Ng|\phi|^2 {Q} $ &
${ Q} Ng \phi^2 { Q^*} $ \\
& \\
$-{ Q^*} Ng (\phi^*)^2 { Q}$ & $-\left[ H_{\rm gp}^*
-\mu + { Q^*} Ng|\phi|^2 { Q^*}\right]$
\end{tabular}\right)
\label{eq:opL}
\end{equation}
\end{center}
where
\begin{equation}
H_{\rm gp} = h_1 + N g |\phi|^2
\label{eq:Hgp}
\end{equation}
and $Q$ projects orthogonally to $\phi$:
\begin{equation}
Q=1-|\phi\rangle \langle\phi|\,.
\end{equation}
The complex conjugation indicated by the star is taken in the $r$-basis so that
e.g.\ $Q^*$ projects orthogonally to the state with wavefunction $\phi^*$.
The projection operator appears as well in the commutation relations of
$\hat{\Lambda}$ and $\hat{\Lambda}^\dagger$:
\begin{equation}
[\hat{\Lambda}(\vec{r}\,),\hat{\Lambda}^\dagger(\vec{s}\,)] =
	\langle \vec{r} \, |Q| \vec{s} \, \rangle \;.
\label{eq:commlambda}
\end{equation}

The linear equations of motion (\ref{eq:evollambda}) for $\hat{\Lambda}$ correspond to an approximation
of the Hamiltonian which is quadratic in $\hat{\Lambda}$. This quadratic approximation can be obtained
by inserting the decomposition (\ref{eq:psi}) in the Hamiltonian, by neglecting terms
cubic or quartic in $\delta \hat{\psi}$, and by using:
\begin{eqnarray}
a_{\phi_{\rm ex}}^\dagger a_{\phi_{\rm ex}} &=& \hat{N} - \delta\hat{N} \\
\delta\hat{N} &\equiv& \int \delta \hat{\psi}^\dagger \delta \hat{\psi}
	\simeq  \int \hat{\Lambda}^\dagger \hat{\Lambda} \,.
\label{eq:def_dN}
\end{eqnarray} 
As equation (\ref{eq:GPE}) is satisfied,
no term linear in $\hat{\Lambda}$ is left in the Hamiltonian. 
One is left with the quadratic Hamiltonian \cite{Houches}:
\begin{equation}
\hat{H}_{\rm quad}=f(\hat{N}) + \int
\frac{1}{2} (\hat{\Lambda}^\dagger,-\hat{\Lambda}) \cdot
{ {\cal L}} \left(
\begin{tabular}{c} $\hat{\Lambda}$ \\ $\hat{\Lambda}^\dagger$
\end{tabular}\right)
\label{eq:hamlambda}
\end{equation}

At thermal equilibrium in the canonical ensemble, the non condensed atoms are then described 
by the density operator:
\begin{equation}
\hat{\sigma}_{\rm nc} = \frac{1}{Z} e^{-\beta \hat{H}_{\rm quad}} \,.
\label{eq:sigma}
\end{equation}
The usual way of proceeding at this point is to diagonalize the operator ${\cal L}$. 
Let us introduce the eigenvalues and eigenvectors of ${\cal L}$:
\begin{equation}
{\cal L} \left( \begin{tabular}{c} $u_k$ \\ $v_k$ \end{tabular}\right) =
\epsilon_k  \left( \begin{tabular}{c} $u_k$ \\ $v_k$ \end{tabular}\right) \,.
\end{equation}
We suppose that all the eigenvalues $\epsilon_k$ are real so that the equilibrium 
condensate wavefunction is a dynamically stable solution of the Gross-Pitaevskii 
equation. Moreover we restrict the notation $(u_k,v_k)$  
to the modes satisfying the normalization and orthogonality conditions: 
\begin{equation}
\int u_k^* u_{k^\prime} - v_k^* v_{k^\prime} =\delta_{k,k^\prime} \;.
\label{eq:othonorm_uv}
\end{equation}
To each mode $(u_k,v_k)$ is associated a ``dual mode''  $(v_k^*,u_k^*)$, 
of energy $-\epsilon_k$ and satisfying obviously 
$\int  v_k^* v_{k^\prime} - u_k^* u_{k^\prime} =-\delta_{k,k^\prime}$.
In general the $(u_k,v_k)$'s plus the dual modes form a complete family in the subspace 
orthogonal to $(\phi,0)$ and $(0,\phi^*)$ so that
the field $\Lambda$ can then be expressed as:
\begin{equation}
\left( \begin{tabular}{c} $\hat{\Lambda}(\vec{r}\,)$ \\ 
	$\hat{\Lambda}^\dagger(\vec{r}\,)$ \end{tabular} \right) =
\sum_k \; \hat{b}_k \left( 
\begin{tabular}{c} $u_k(\vec{r}\,)$ \\ $v_k(\vec{r}\,)$ \end{tabular}\right) +
\hat{b}^\dagger_k \left( 
\begin{tabular}{c} $v^*_k(\vec{r}\,)$ \\ $u^*_k(\vec{r}\,)$ \end{tabular}\right)
\label{eq:explambda}\,.
\end{equation}
The normalization condition of the $(u_k,v_k)$'s ensures that the $b_k,b_k^\dagger$
satisfy bosonic commutation relations.
If (\ref{eq:explambda}) is injected in (\ref{eq:hamlambda}), the quadratic 
Hamiltonian takes the form:
\begin{equation}
H_{\rm quad} = E_0 +
        \sum_k  \epsilon_k \; \hat{b}^\dagger_k  \hat{b}_k
\label{eq:hquadrdiag}
\end{equation}
corresponding to a set of decoupled harmonic oscillators. For thermodynamical
stability of the condensate we must require that all the $\epsilon_k$ are positive.
Physical quantities can then be readily extracted from $\hat{\sigma}_{\rm nc}$.
The inconvenient of this method is that the diagonalization
of ${\cal L}$ in two and three dimensions is a very heavy numerical task in the
absence of particular symmetries of the problem.

\section{Formulation of the Bogolubov theory in terms of the Wigner function}
\label{sec:wigner}

We wish to deal with complex numbers and complex functions instead of operators and fields
\[ 
\hat{b}^\dagger_k \;\;,\;\; \hat{b}_k  \hspace{1.5cm} {\bf \rightarrow}  
				\hspace{1.5cm} b^*_k \;\;,\;\; b_k
\]
\[ 
\hat{\Lambda}^\dagger(\vec{r}\,) \;\;,\;\; 
	\hat{\Lambda}(\vec{r}\,)  \hspace{1.5cm} {\bf \rightarrow}  
	\hspace{1.5cm} \Lambda^*(\vec{r}\,) \;\;,\;\; \Lambda(\vec{r}\,) \;.
\]
For this reason we introduce the Wigner quasi distribution function 
$W(\{b_k\},\{b_k^*\})$ defined as the Fourier transform of the characteristic
function $\chi(\{\alpha_k\})$: 
\begin{equation}
\chi(\{\alpha_k\},\{\alpha_k^*\}) = \mbox{Tr} \left[\hat{\sigma}_{\rm nc}  
	\prod_k \exp{(\alpha_k \hat{b}_k^\dagger - {\alpha_k^*} \hat{b_k} )} \right]
\end{equation}
and
\begin{equation}
W(\{b_k\},\{b_k^*\})= \int \chi(\{\alpha_k\},\{\alpha_k^*\}) \;
	\prod_k \exp \left[ (\alpha_k^* b_k - \alpha_k b_k^*)\right] \;
	\frac{(d \Re e \, \alpha_k) (d \Im m \, \alpha_k)}{\pi} \;.
\end{equation}
With the diagonalized form of the quadratic Hamiltonian (\ref{eq:hquadrdiag}), 
the density operator (\ref{eq:sigma}) corresponds to decoupled harmonic 
oscillators at thermal equilibrium, so that:
\begin{equation}
W(\{b_k\},\{b_k^*\})= \prod_k  2 \tanh \left( \frac{\beta \epsilon_k}{2} \right)
	\exp\left[-2 |b_k|^2 \mbox{tanh} (\beta \epsilon_k /2)\right] \;.
\label{eq:wignerdiag}
\end{equation}
In the high temperature limit $k_B T \gg \epsilon_k$ the Wigner distribution for mode 
$k$ is simply the Boltzmann distribution for a classical field. 
In the low temperature limit $k_B T \ll \epsilon_k$ it is worth noting that 
the Wigner distribution has a finite width tending to 1/2: even at zero temperature, the
c-number quantities $b_k$ fluctuate mimicking the quantum fluctuations of the 
operators $\hat{b}_k$. 

Since we want to avoid the diagonalization of the operator ${\cal L}$, we wish now 
to express the Wigner function (\ref{eq:wignerdiag}) in terms of the functions 
$\Lambda$ and $\Lambda^*$ corresponding to the field operators $\hat{\Lambda}$ and
$\hat{\Lambda}^\dagger$:
\begin{equation}
\left( \begin{tabular}{c} ${\Lambda}(\vec{r}\,)$ \\ 
	${\Lambda}^*(\vec{r}\,)$ \end{tabular} \right) \equiv
\sum_k \; {b}_k \left( 
\begin{tabular}{c} $u_k(\vec{r}\,)$ \\ $v_k(\vec{r}\,)$ \end{tabular}\right) +
 b^*_k \left( 
\begin{tabular}{c} $v^*_k(\vec{r}\,)$ \\ $u^*_k(\vec{r}\,)$ \end{tabular}\right) \,.
\label{eq:lambda-b}
\end{equation}
By using (\ref{eq:lambda-b}) and the properties (\ref{eq:othonorm_uv})
of the $u_k(\vec{r}\,)$ and $v_k(\vec{r}\,)$,
one can show that the Wigner function takes the form:
\begin{equation}
W(\Lambda,\Lambda^*) =  \mbox{det}\left[4 \eta \tanh \frac{\beta {\cal L}}{2} \right]^{1/2}
\exp\left\{- \int (\Lambda^*,{\Lambda}) \cdot \;\eta \,\tanh\frac{\beta{\cal L}}{2}
\;\left(\begin{tabular}{c}
${\Lambda}$ \\  ${\Lambda}^*$ 
\end{tabular}\right)\right\}
\label{eq:wignerlambda}
\end{equation}
where $\eta$ is the matrix:
\begin{equation}
\eta = \left(\begin{tabular}{cc} {\rm Id} & 0 \\ 0 & $-{\rm Id}$
\end{tabular}\right) \,,
\end{equation}
and the determinant is defined in the space orthogonal to $(\phi,0)$ and $(0,\phi^*)$.

The quantum mechanical mean values of totally symmetrized operators are then
simply given as averages over the Wigner distribution \cite{WallsQOPT}; 
e.g.:
\begin{equation}
\langle \Lambda(\vec{r_1}\,) \Lambda^*(\vec{r_2}\,) \rangle_W = \frac{1}{2}
\langle \hat{\Lambda}^\dagger(\vec{r_1}\,) \hat{\Lambda}(\vec{r_2}\,)
 +\hat{\Lambda}(\vec{r_2}\,) \hat{\Lambda}^\dagger(\vec{r_1}\,) \rangle.
\label{eq:avedens_nc}
\end{equation}
Note that  according to (\ref{eq:commlambda}) 
the previous expression gives infinity for $\vec{r_1}=\vec{r_2}$! 
In practice we will not encounter this problem because we will work on a grid
with a finite number of modes.

\section{Brownian motion simulation}
\label{sec:simul}

Since we quadratized the Hamiltonian with respect to $\hat{\Lambda}$
and $\hat{\Lambda}^\dagger$, the Wigner distribution (\ref{eq:wignerlambda})
is Gaussian and can be recovered as the steady-state probability distribution
of a conveniently chosen Brownian motion of the functions $\Lambda$ and $\Lambda^*$.
In this section we show that one can find such Brownian motion in a convenient way for 
numerical simulations.
First of all we discretize our problem on a grid. To this aim we introduce
the vector $\vec{\Lambda}$ representing the set of values $\{\Lambda({\vec r_l}\,)\}$ where
$\vec{r_l}=(x_i,y_j,z_k)$ are points of the grid:
\begin{equation}
({\vec \Lambda})_l \equiv \Lambda(\vec{r_l}\,) \;.
\end{equation}
If $dV$ is the volume element of the grid, the Wigner function (\ref{eq:wignerlambda}) 
takes the form:
\begin{equation}
W(\Lambda,\Lambda^*) = {\cal A}  \exp\left[-dV \;
(\vec{\Lambda}^*,\vec{\Lambda})
\cdot \;\eta \,\tanh\frac{\beta{\cal L}}{2}
\;\left(\begin{tabular}{c}
$\vec{\Lambda}$ \\  $\vec{\Lambda}^*$
\end{tabular}\right)\right]
\end{equation}
where now ${\cal L}$ is represented by a matrix, and
where we did not write explicitly the normalization factor.
The vector $\vec{\Lambda}$ is then submitted to the following stochastic
evolution during the time step $dt$:
\begin{equation}
d\vec{\Lambda} =  { \vec{F}}dt
+ { Y_{11}} \, d\vec{\xi} + { Y_{12}} \, d\vec{\xi}^*
\label{eq:stoch}
\end{equation}
where ${\vec F}$ is a linear friction force 
\begin{equation}
{ \vec{F}}=- { \alpha_{11}} \;  
	\vec{\Lambda } - { \alpha_{12}} \;\vec{\Lambda}^*
\label{eq:force}
\end{equation}
and $d \vec{\xi}$ is a noise term satisfying:
\begin{eqnarray}
\overline{ d\xi(\vec{r}\,) } & = & 0 \\
\overline{ d\xi(\vec{r}_1) d\xi^*(\vec{r}_2) } &=& 2\frac{dt}{dV}\;
\left[\delta_{\vec{r}_1,\vec{r}_2} - dV \, 
\phi(\vec{r}_1) \phi^*(\vec{r}_2)\right] \label{eq:noiseortho} \\
\overline{ d\xi(\vec{r}_1) d\xi(\vec{r}_2) }  &=& 0 \,,
\end{eqnarray}
where the overline denotes the average over the time interval $dt$.
In (\ref{eq:stoch}) and (\ref{eq:force}), 
$\vec{\Lambda}$ and $\vec{\xi}$ are vectors
whose length ${\cal N}$ is the size of the grid, while $\alpha_{ij}$ and $Y_{ij}$
$(i,j=1,2)$ are ${\cal N} \times {\cal N}$ matrices. 
The prescription (\ref{eq:noiseortho}) is equivalent to the usual prescription
for spatially $\delta$-correlated Gaussian white noise 
with the additional requirement that the
noise should be orthogonal to the condensate wavefunction $\phi$,  
so that $\Lambda$ remains within the subspace orthogonal to $\phi$.
It is important to note that the choice of the `time' variable $t$ in the Brownian
motion simulation is totally arbitrary; in what follows $dt$ will have the dimension
of an energy.

Let $P(\vec{\Lambda},\vec{\Lambda}^*,t)$ be the probability distribution
of $\vec{\Lambda}$ and $\vec{\Lambda}^*$ at time $t$.
The stochastic equations (\ref{eq:stoch}) can then be turned into a Fokker-Plank
equation for $P(\vec{\Lambda},\vec{\Lambda^*},t)$:
\begin{equation}
\partial_t P+\sum_{\vec{r}} \left[
\partial_{\Lambda(\vec{r}\,)} J(\vec{r}\,) +\mbox{c.c.}
\right]=0
\label{eq:FPE}
\end{equation}
where $J(\vec{r}\,)$ is a density current:
\begin{equation}
J(\vec{r}\,) = { F(\vec{r}\,)} P -\frac{1}{dV}\sum_{\vec{s}} \partial_{\Lambda(\vec{s}\,)}
({ D_{12}(\vec{r},\vec{s}\,)}P) + \partial_{\Lambda^*(\vec{s}\,)}
({ D_{11}(\vec{r},\vec{s}\,)}P) \,
\end{equation}
and $D_{ij}$ are constant diffusion matrices: 
\begin{eqnarray}
(D_{11})(\vec{r},\vec{s}\,) &=& 
\frac{dV}{2 dt}  \; \overline{ d\Lambda(\vec{r}\,)
d\Lambda^*(\vec{s}\,) }
\label{eq:diff1}\\
(D_{12})(\vec{r},\vec{s}\,) &=& 
\frac{dV}{2 dt} \; \overline{ d\Lambda(\vec{r}\,)
d\Lambda(\vec{s}\,) }
\,. \label{eq:diff2} 
\end{eqnarray}
Neglecting terms on the order of $dt$, we can express the 
identities (\ref{eq:diff1}-\ref{eq:diff2}) more synthetically 
in terms of a diffusion matrix:
\begin{equation}
{ D} \equiv \left(\begin{tabular}{cc}$D_{11}$ & $D_{12}$ \\
$D_{12}^*$ & $D_{11}^*$\end{tabular}\right)= { Y Y^\dagger} 
\end{equation}
where we have introduced the noise matrix
\begin{equation}
\label{eq:matrice_Y}
Y = \left(\begin{tabular}{cc}$Y_{11}$ & $Y_{12}$ \\
$Y_{12}^*$ & $Y_{11}^*$\end{tabular}\right) \;.
\end{equation}

The steady-state solution of (\ref{eq:FPE}) with $J(\vec{r}\,)=0$ 
is found to be:
\begin{equation}
P \propto  \exp\left[-\frac{dV}{2}(\vec{\Lambda}^*,\vec{\Lambda}) \cdot
D^{-1}\alpha\left(
\begin{tabular}{c} $\vec{\Lambda}$ \\ $\vec{\Lambda}^*$ \end{tabular}
\right)\right]
\label{eq:ststate}
\end{equation}
with the friction matrix $\alpha$ defined as:
\begin{equation}
\label{eq:matrice_alpha}
\alpha = \left(\begin{tabular}{cc}$\alpha_{11}$ & $\alpha_{12}$ \\
$\alpha_{12}^*$ & $\alpha_{11}^*$\end{tabular}\right) \;.
\end{equation}
If we now require that the steady-state distribution (\ref{eq:ststate})
coincides with the Wigner distribution (\ref{eq:wignerlambda}) describing 
thermal equilibrium, we obtain:
\begin{equation}
\alpha = 2 \,  D\, \eta \tanh\left(\frac{\beta {\cal L}}{2}\right) \,.
\label{eq:einstein}
\end{equation}
Equation (\ref{eq:einstein}) is simply the generalization to a field of
Einstein's relation for Brownian motion of a particle: in the high temperature limit 
the diffusion term divided by the friction term is proportional to temperature.

Provided that the Einstein's relation (\ref{eq:einstein}) is satisfied,
the choice of $\alpha$ and $Y$ (which determines $D$) is not unique. 
It turns out that a convenient choice of the friction and 
noise terms from the point of view of a numerical simulation is given by: 
\begin{equation}
\alpha=\left[ \cosh \left( \frac{\beta {\cal L}}{2} \right) \right] \,
\, \eta \, \left[ \sinh \left( \frac{\beta {\cal L}}{2} \right) \right]/
	\left(\frac{\beta}{2}\right)
\label{eq:fric}
\end{equation}
and:
\begin{equation}
Y=\left[ \cosh \left( \frac{\beta {\cal L}}{2} \right) \right] \frac{1}{\beta^{1/2}}\,.
\label{eq:noise}
\end{equation}
One can check that the choice (\ref{eq:fric}) and (\ref{eq:noise})
satisfies (\ref{eq:einstein}) by using the following properties of the
operators ${\cal L}$ and $\eta$:
\begin{equation}
\eta \, {\cal L}\,  \eta = {\cal L}^\dagger  \hspace{2cm} \mbox{and} 
\hspace{2cm} \eta^{-1} =  \eta  \,.
\end{equation}
One also has to check that the choice (\ref{eq:fric}) and (\ref{eq:noise})
reproduces the structure of the matrices (\ref{eq:matrice_Y}) and
(\ref{eq:matrice_alpha}), that is the second line of the matrix 
is obtained from the first line by complex 
conjugation and exchange of indices 1 and 2, a property that can be
formalized as
\begin{equation}
S\, \alpha S\, = \alpha^*  \hspace{2cm} \mbox{and}
\hspace{2cm} S\, Y S\, = Y
\end{equation}
where $S$ is the matrix
\begin{equation}
S = S^{-1} = \left(\begin{tabular}{cc} 0 & {\rm Id} \\  ${\rm Id}$ & 0
\end{tabular}\right) \,.
\end{equation}
From the properties $S\, {\cal L} \, S = -{\cal L}^*$ and
$S\, \eta \, S = - \eta$ one finds 
that (\ref{eq:fric}) and (\ref{eq:noise}) have the right structure indeed.

The convenience of the choice (\ref{eq:fric}) and (\ref{eq:noise})
relies in the fact that one then is brought to calculate 
the action of the operators $\exp\left( \frac{\beta {\cal L}}{2} \right)$ 
on the functions $\Lambda$, which is easily implemented by using 
a splitting plus Fourier transform technique (more details about the numerical
simulation will be given in the end of the next section).

\section{Test of the method for a 1D trapped Bose gas}
\label{sec:resul}
\begin{figure}[htb]
\centerline{\epsfxsize=14cm \epsfbox{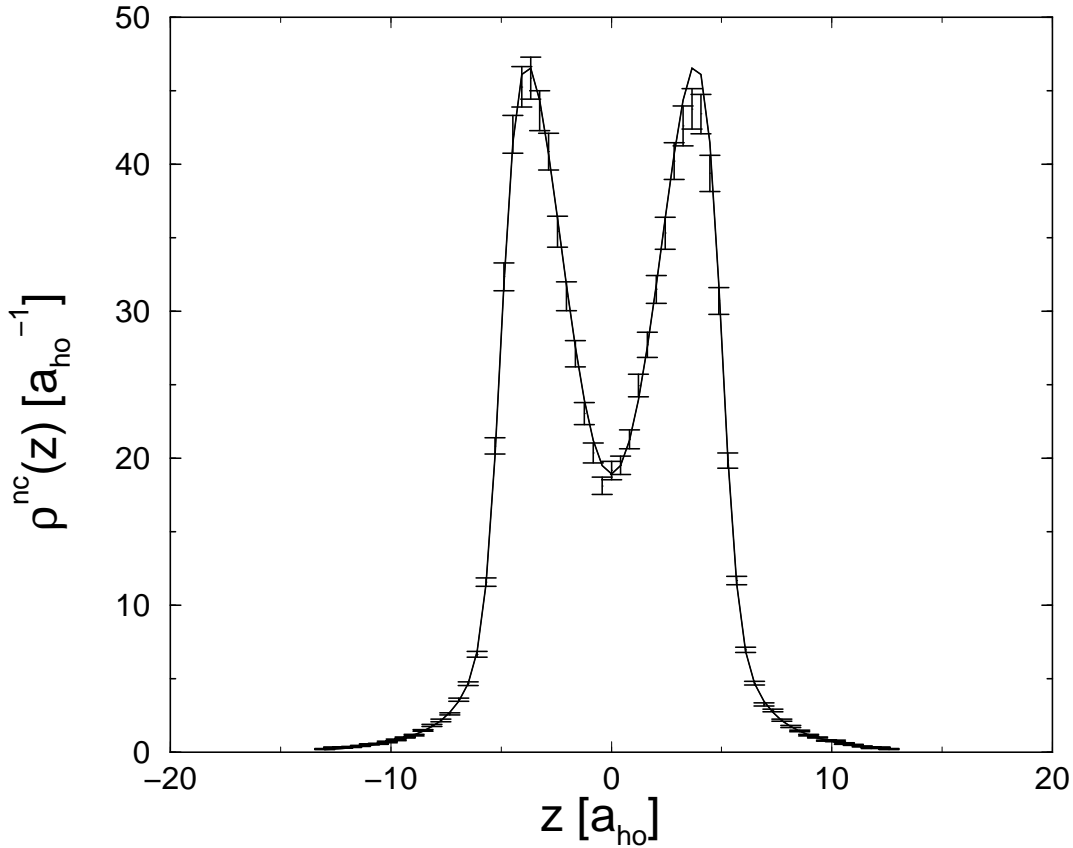}}
\caption{Spatial density of non condensed atoms in presence of
a condensate in a 1D harmonic trap, for: $N=10^4$,
$k_BT=30 \hbar \omega$, $g_{1D}=0.01 \hbar (\omega/m)^{1/2}$. The  
corresponding chemical potential is $\mu=14.1 \hbar \omega$.
Full line: direct diagonalization.
Points with error bars: Monte Carlo simulation with 2000 realizations.
The length scale used is the harmonic oscillator length $a_{\rm ho}=\sqrt{\hbar/m\omega}$.}
\label{fig:dens_nc}
\end{figure}

As a test of the method we have made a numerical comparison between the Monte Carlo
simulation and the direct diagonalization of the operator ${\cal L}$ for the case of a
Bose-Einstein condensate in a harmonic potential with frequency $\omega$ in one spatial dimension.
Physically this situation would correspond to a very elongated cigar shaped trap
with a longitudinal trap frequency $\omega$ and a strong transversal confinement 
$\hbar \omega_{\perp} > k_B T$, 
so that most of the atoms are in the transverse ground state $\phi_\perp$ of the trap. 
In these conditions the field operator can be approximated by:
\begin{equation}
\hat{\psi}(\vec{r}\,)=\phi_\perp(x,y)  \, \hat{\psi}(z) 
\end{equation}
where the operators $\hat{\psi}^\dagger{(z)}$ and $\hat{\psi}{(z)}$ have the usual bosonic
commutation relations for field operators in one dimension.
We have then reduced our problem to one spatial dimension with 
an effective coupling constant $g_{1D}$ \cite{Maxim}:
\begin{equation}
g_{1D}=g \, \int dx \int dy |\phi_\perp(x,y)|^4 \,.
\end{equation}
We wish to calculate the spatial density of non condensed atoms.
In order to calculate the average 
$\langle \hat{\Lambda}^\dagger(z) \hat{\Lambda}({z}) \rangle$
we use the formula (\ref{eq:avedens_nc}) and equation (\ref{eq:commlambda})
in its discretized version, to obtain:
\begin{equation}
\langle \hat{\Lambda}^\dagger(z) \hat{\Lambda}(z) \rangle =
\langle {\Lambda}^*(z) {\Lambda}(z) \rangle - \frac{1}{2} 
	\left[ \frac{1}{dz} - |\phi(z)|^2  \right] \,.
\label{eq:disc_ver}
\end{equation}
The result for the non condensed cloud spatial density as a function of $z$
is reported in figure \ref{fig:dens_nc}. We have chosen a total number of atoms 
$N=10^4$, a temperature $k_BT=30 \hbar \omega$, and a coupling constant 
$g_{1D}=0.01 \hbar (\hbar\omega/m)^{1/2}$ leading to a chemical potential $\mu=14.1 \hbar \omega$. 
The full line is the direct diagonalization result
and the points with the error bars are obtained with our Monte Carlo
simulation with 2000 realizations. 
The non condensed cloud is repelled by the condensate which is sitting
in the center of the harmonic trap, which explains the double peak structure 
of figure \ref{fig:dens_nc}. 

We can also get information on the number of condensate particles 
by calculating the number of non-condensed particles, since the total
number of particles is fixed. The mean number of non-condensed particles
is given by the spatial integral of Eq.(\ref{eq:disc_ver}):
\begin{equation}
\langle \delta \hat{N}\rangle = \langle\delta N\rangle_{W}
-\frac{{\cal N}-1}{2}
\end{equation}
where the operator $\delta\hat{N}$, defined in (\ref{eq:def_dN}),
is replaced at the present order by its approximation in terms
of $\hat{\Lambda}$,
$\cal N$ is the number of modes in the simulation (that is the
number of points on the spatial grid) and
\begin{equation}
\delta N \equiv dz\, \sum_{z} \Lambda^*(z)\Lambda(z).
\end{equation}
After some algebra, involving the total symmetrization of
a product of 4 operators $\hat{\Lambda}$ or $\hat{\Lambda}^\dagger$,
we also obtain
\begin{equation}
\langle \delta\hat{N}^2\rangle - \langle \delta\hat{N}\rangle ^2 =
\langle \delta N^2\rangle_{W} -\langle\delta N\rangle_W^2 -
\frac{{\cal N}-1}{4}.
\end{equation}
For the parameters of the figure we get the mean values and the
standard deviations
\begin{eqnarray}
(\langle \delta \hat{N}\rangle)_{\rm MC} = 385 \pm 6 \hspace{2cm}&\mbox{and}& \hspace{2cm} 
(\langle\delta \hat{N}\rangle)_{\rm diag} = 391 \,, \\
\sigma(\delta \hat{N})_{\rm MC} = 269 \pm 10 \hspace{2cm}&\mbox{and}& \hspace{2cm} 
\sigma(\delta \hat{N})_{\rm diag} = 279 \,, \\
\end{eqnarray}
where the subscript \lq MC' represents the results of the Monte Carlo
simulation and \lq diag' indicates the predictions of a direct
diagonalization of ${\cal L}$.

We give in the following some details and tricks about the numerical simulation.
The first thing do to is to solve the steady-state Gross-Pitaevskii equation 
to find the function $\phi$. This is done by imaginary time evolution.
In practice we iterate the action of the evolution operator
$\exp\left(- d\tau \, H_{\rm gp} \right)$ (where $H_{\rm gp}$ is given in (\ref{eq:Hgp}))
on an initial wavefunction, renormalizing the wavefunction and  calculating $\mu$ at each 
time step $d\tau$, until $\mu$ reaches a stationary state within the required accuracy. 
If $d\tau$ is small enough, we can approximate the evolution of the wavefunction
in each interval $d\tau$ by splitting the exponential 
$\exp\left(-d\tau \, H_{\rm gp} \right)$ into the exponential of kinetic energy and
the exponential of potential energy:
\begin{equation}
\phi(\tau+d\tau)=\exp\left( -d\tau \, H_{\rm gp} \right) \phi(\tau) \simeq 
	\exp\left( -d\tau \, E_{\rm cin} \right)
        \exp\left( -d\tau \, E_{\rm pot} \right) \phi(\tau)      
\label{eq:split}
\end{equation}
where:
\begin{eqnarray}
E_{\rm cin}&=& \frac{{\vec p\,}^2}{2m} \\
E_{\rm pot}&=& U(z)+ Ng |\phi(z,\tau)|^2 \,.
\end{eqnarray}
The advantage of the splitting in (\ref{eq:split}) is that the action
of $E_{\rm cin}$ and $E_{\rm pot}$ on a vector can be easily calculated by going back 
and forth in the Fourier space and in the real space. 
This procedure is called the splitting method and is it a common technique to
solve the steady-state Gross-Pitaevskii equation. Let us now consider the
Brownian motion simulation: because of our choice
(\ref{eq:fric}) and (\ref{eq:noise}) we have to calculate the action of
the exponential of $\beta {\cal L}/2$ over a vector. It is convenient to write:
\begin{equation}
\exp\left(\frac{\beta}{2} {\cal L}\right) = 
	\left[ \exp\left(d\tau {\cal L}\right) \right]^n
	\hspace{1cm} \mbox{with} \hspace{1cm}  d\tau = \frac{\beta/2}{n} \,.
\label{eq:chop}
\end{equation}
The idea underlying equation (\ref{eq:chop}) is to choose $n$ large enough so that the 
splitting approximation sketched above can be applied.
Before implementing the splitting however, we still have to perform
some transformation.
We notice that if acting on functions orthogonal to the condensate wavefunctions, 
$\exp\left(d\tau {\cal L}\right)$ can be expressed as:
\begin{equation}
\exp\left(d\tau {\cal L}\right) =  
	\left(\begin{tabular}{cc} $Q$ & 0 \\ 0 & $Q^*$ 
	\end{tabular}\right)	\exp\left(d\tau {\cal L}_{gp} \right) 
\end{equation}
where ${\cal L}_{gp}$ is the operator obtained by linearization of the Gross-Pitaevskii
equation, which is the same as ${\cal L}$ given by (\ref{eq:opL}) without the projectors.
We are then led to the action of $\exp\left(d\tau {\cal L}_{gp} \right)$ on a vector and
will do it approximately by using the splitting technique. In order to be consistent
however we must take care that in our simulation the approximate action
of ${\cal L}_{gp}$ corresponds exactly to the linearization of 
the approximate Gross-Pitaevskii equation in imaginary time (\ref{eq:split})
that we have used to find the condensate wavefunction. We then obtain:
\begin{equation}
\exp\left(d\tau {\cal L}_{gp} \right) \simeq 
	\exp\left( -d\tau E_{\rm cin} \right) 
	\exp\left[ -d\tau (E_{\rm pot}-\mu) \right] \;
	\exp\left[ N g \, d\tau \left(\begin{tabular}{cc} $|\phi|^2$ & $\phi^2$ \\ 
      	$-{\phi^*}^2$ & $-|\phi|^2$ \end{tabular}\right) \right] \,,
\end{equation}  
where the matrix in the last factor has the nice property of giving zero if squared.

A last point concerns the choice of the time step $dt$ of the Brownian 
motion, and of the total simulation time $t_{\rm max}$ 
guaranteeing that the steady-state is reached. 

The time step $dt$ should satisfy $\alpha_{\rm max} dt\ll 1$ where
$\alpha_{\rm max}$ is the largest eigenvalue of the friction matrix
Eq.(\ref{eq:fric}). To estimate $\alpha_{\rm max}$ we consider the eigenmode
$(u_{\rm max},v_{\rm max})$
of $\cal L$ with the largest eigenvalue $e_{\rm max}$.
This value is well above the chemical potential $\mu$, so that $v_{\rm max}$ can be
neglected as compared to $u_{\rm max}$, in the so-called Hatree-Fock approximation.
In this case this largest energy eigenmode is an approximate eigenvector of
the friction matrix, and we get the condition
\begin{equation}
\alpha_{\rm max}dt\simeq dt\sinh(\beta e_{\rm max})/\beta\ll 1.
\end{equation}
In a one-dimensional harmonic trap one expects $e_{\rm max}\simeq  e_{\rm max}^{\rm ho}
-\mu$ in the Hartree-Fock approximation, where $e_{\rm max}^{\rm ho}$ is the maximal
energy level of the bare trapping potential.

The condition to be satisfied by the duration $t_{\rm max}$ of the simulation
is $\alpha_{\rm min} t_{\rm max} \gg 1$ where $\alpha_{\rm min}$ is the
smallest eigenvalue of the friction matrix. We assume here that the corresponding eigenmode of
$\alpha$ has components on energy modes of $\cal L$ with energy much smaller than $k_B T$
only. In this case $\alpha_{\rm min}$ 
corresponds approximately to the minimum eigenvalue of $\eta {\cal L}$. 
In the case of real condensate wavefunction, the determination of the 
minimum eigenvalue of $\eta {\cal L}$ orthogonal to the condensate wavefunction
boils down to the calculation of the first excited state energy of $H_{gp} -\mu$
\cite{Houches};
following \cite{SinatraGora} we then find the condition:
\begin{equation}
t_{\rm max} \; \frac{\hbar^2}{2m} \left( \int |\phi|^2 z^2 \right)^{-1} \gg 1 \,.
\end{equation}
In the weakly interacting regime one finds $t_{\rm max} \gg \hbar\omega$.
In the opposite Thomas-Fermi regime one finds that $t_{\rm max}$ scales as $\mu\gg\hbar\omega$.
In practice, in the case of figure \ref{fig:dens_nc} we had:
$dt~=~1.15\times 10^{-4}/\hbar \omega$ and $t_{\rm max}~=~56.9/\hbar \omega$.

\section{The time-Dependent Case}
\label{sec:timedep}

Let us consider the following situation: initially we have a Bose-Einstein
condensate in a thermodynamically stable state at thermal equilibrium. 
Suddenly a parameter of the system is changed
(e.g. the trapping potential or the scattering length) so that the system
is no more in an equilibrium state and it undergoes a dynamical evolution
during a given time, after which we perform a measurement.

In order to study this situation we need to recall some more results
of the time-dependent number conserving Bogolubov theory of \cite{CastinDum}.
It is shown in this paper that the expectation value
of a one-particle observable $\hat{X}=\sum_{i=1}^N X(i)$ is given by:
\begin{eqnarray}
\langle X \rangle &\simeq& (N-\langle \delta \hat{N} \rangle)
	\langle \phi |X(1)|\phi \rangle +
	\langle \phi |X(1)|\phi_{(2)} \rangle +
	\langle \phi_{(2)} |X(1)|\phi \rangle  \nonumber \\
	& & + \int d{\vec{r}} \int d{\vec{r^\prime}} 
	\langle {\vec{r}} \, |X(1)|{\vec{r^\prime}} \rangle \,
	\langle \hat{\Lambda}^\dagger({\vec{r}}\,) 
		\hat{\Lambda}({\vec{r^\prime}}) \rangle + 
	O\left(\frac{1}{\sqrt{N}}\right)\;.
\label{eq:expX}
\end{eqnarray}
In this equation the first term contains the leading contribution to
$\langle X \rangle$ scaling as $N$ in the limit (\ref{eq:tdlimit}), 
where $\phi$ is solution of the Gross-Pitaevskii equation (\ref{eq:GPE}), and
\begin{equation}
\langle \delta \hat{N} \rangle = \int d{\vec{r}} \;
 \langle \hat{\Lambda}^\dagger({\vec{r}},t)
                 \hat{\Lambda}({\vec{r}},t) \rangle \;,
\end{equation} 
so that $N-\langle \delta \hat{N} \rangle$ is the mean number of particles
in the condensate. 
The other terms involve the first correction to condensate
wavefunction $\phi_{(2)}$ and the contribution
of the non condensed atoms to the expectation value of the observable. 
The equation of motion of $\phi_{(2)}$ given in \cite{CastinDum} is:
\begin{equation}
 \left( i \hbar \frac{d}{dt} - {\cal L}(t) \right)
\left( \begin{array}{r}
\phi_{(2)}(t) \\
\phi_{(2)}^*(t)
\end{array} \right) 
= 
\left( \begin{array}{r}
Q(t)R(t) \\
-Q^*(t)R^*(t)
\end{array} \right) 
\label{eq:evolpsi2}
\end{equation}
where ${\cal L}$ is given by Eq.(\ref{eq:opL}) and
\begin{eqnarray}
\label{SOURCE}
R(\vec{r}\,) &=& -g N 
{| \, {\phi({\vec{r}\,})} \, |}^2 \phi({\vec{r}\,}) 
\langle 1+ \int d\vec{s}\, {\hat{\Lambda}}^\dagger(\vec{s}\,) {\hat{\Lambda}}(\vec{s}\,)\rangle 
\nonumber \\
&+& 2 g N \phi({\vec{r}\,}) 
\langle {\hat{\Lambda}}^\dagger({\vec{r}\,}){\hat{\Lambda}}({\vec{r}\,})\rangle 
+g N \phi^*({\vec{r}\,})
\langle {\hat{\Lambda}}({\vec{r}\,}){\hat{\Lambda}}({\vec{r}\,}) \rangle 
\nonumber \\
&-& g N \int d\vec{s}\, {| \, {\phi(\vec{s}\,)} \, |}^2 
 \langle \left[ {\hat{\Lambda}}^\dagger(\vec{s}\,)\phi(\vec{s}\,)+ 
  {\hat{\Lambda}}(\vec{s}\,)\phi^*(\vec{s}\,) \right] 
  {\hat{\Lambda}}({\vec{r}\,})\rangle \;.
\end{eqnarray}
The first term in Eq.(\ref{SOURCE})
corrects the overestimation of the number of condensed particles in calculating
their mutual interaction
($N \rightarrow N-(1+\langle \delta\hat{N}\rangle$) in the Gross-Pitaevskii equation.
The terms in the second line
describe the interaction of the condensed particles and the 
non condensed ones. We refer to \cite{Houches} and 
\cite{CastinDum} for more details.

The implementation of our method for the time-dependent case then consists
of three parts: (i) we generate many stochastic realizations of the initial state 
sampling the correct equilibrium distribution. (ii) We perform a 
deterministic evolution in real time of the stochastic initial states. (iii)
We calculate the expectation value by averaging over the stochastic realizations. 

Let us first concentrate on part (i): $\phi$ which is the condensate wavefunction 
to the lowest approximation, is found by solving the Gross-Pitaevskii equation 
(\ref{eq:GPE}) in steady-state.
Once we have $\phi$, we can use the Monte Carlo simulation described in section 
\ref{sec:simul} to generate many realizations of the stochastic functions 
$\Lambda$ and $\Lambda^*$ representing the non condensed fraction. To calculate
$\phi_{(2)}$, we calculate the source term (\ref{SOURCE}) and
solve (\ref{eq:evolpsi2}) at steady-state with an imaginary time evolution:
\begin{equation}
 \frac{d}{d\tau} 
\left( \begin{array}{r}
\phi_{(2)}(t) \\
\phi_{(2)}^*(t)
\end{array} \right)
=
-\eta {\cal L}(t=0)
\left( \begin{array}{r}
\phi_{(2)}(t) \\
\phi_{(2)}^*(t)
\end{array} \right)                   
-
\left( \begin{array}{r}
Q(t=0)R(t=0) \\
Q^*(t=0)R^*(t=0)
\end{array} \right) \;.
\end{equation}
We note in this respect that the last term in (\ref{SOURCE}) for a given realization
of $\Lambda$ factorizes as $\Lambda(\vec{r}\,)$ times an $\vec{r}\,$-independent
integral over $\vec{s}\,$.

The real time evolution part (ii) is more straightforward since it is deterministic.
The wavefunction $\phi$ evolves according to (\ref{eq:GPE}).
Each stochastic realization of $\Lambda$, $\Lambda^*$ evolves according to
(\ref{eq:evollambda}), and  the corrections $\phi_{(2)}$, $\phi_{(2)}^*$ evolve
according to (\ref{eq:evolpsi2}). The averages in the source term $R$ 
involving $\Lambda$ and $\Lambda^*$ must be calculated at each time step.

\section{Conclusions and Perspectives}
\label{sec:concl}
We have put forward a Monte Carlo method to sample numerically the  
Gaussian density operator for the non condensed modes
obtained in the Bogolubov approach, once it is
expressed in term of the Wigner quasi-distribution function.
This method allows to implement Bogolubov theory both in steady-state and 
in the time-dependent case without having to diagonalize big matrices, which makes 
it readily scalable to two or three spatial dimensions. 
We plan to use this method to study thermodynamics and finite temperature
dynamics of Bose-Einstein condensates with vortices.

In this paper we have sticked to the Bogolubov theory as presented in
\cite{CastinDum}. In this frame, for the time-dependent case,
the Gross-Pitaevskii condensate wavefunction, the non condensed fraction 
and the first correction to the condensate wavefunction must be first obtained in the 
initial equilibrium state and then they must be evolved separately. 

A natural development of the method beyond the Bogolubov approximation 
consists in a Monte Carlo simulation which samples the Wigner distribution 
for the whole field $\psi$ in equilibrium at temperature $T$ followed by the real
time evolution of the field according to the classical field equation:
\begin{equation}
i\hbar \partial_t \psi = h_0 \psi + g |\psi|^2 \psi \,.
\label{eq:class}
\end{equation}
The classical field equation (\ref{eq:class}), which is formally identical
to the Gross-Pitaevskii equation, corresponds in fact to the approximate evolution
in real time of the Wigner quasi distribution function (truncated Wigner) 
\cite{Drummond}.

The classical equations of motion for the field (\ref{eq:class}) has the 
disadvantage of considering only stimulated processes.
Consider for example two colliding condensates in three dimensions,
corresponding to an initial field of the form 
\begin{equation}
\psi(\vec{r}\,) = A\, (\exp(-ikz)+\exp(ikz)) \,.
\end{equation}
By evolving this field according to (\ref{eq:class}) we will populate only modes
that are plane waves of momentum $\hbar n k$, $n$ integer (e.g. $n=2$ corresponds to the
stimulated process $k + k \rightarrow -k + 2k$). In reality we expect to have binary 
collisions producing emerging atoms along any spatial direction. At the level of $\psi$
this would require a broken translational symmetry in the $x$-$y$ transverse plane. 
Fortunately, the advantage of being in the Wigner point of view is that
{\it all the modes} are initially ``filled'' by quantum noise. In this case all
processes can be stimulated and all possible symmetries of $\psi$ are broken.

\medskip
We acknowledge useful discussions with Gora Shlyapnikov, Yuri Kagan and 
Anthony Leggett in the early stage of this work. 
Laboratoire Kastler Brossel is a unit\'e de recherche de 
l'\'Ecole normale sup\'erieure 
et de l'Universit\'e Pierre et Marie Curie, associ\'ee au CNRS.

\end{document}